\documentclass{Interspeech2024}
\usepackage{color}
\usepackage{threeparttable}
\usepackage{soul}
\usepackage{diagbox}

\usepackage{hyperref}
\makeatletter
\def\UrlAlphabet{%
      \do\a\do\b\do\c\do\d\do\e\do\f\do\g\do\h\do\i\do\j%
      \do\k\do\l\do\m\do\n\do\o\do\p\do\q\do\r\do\s\do\t%
      \do\u\do\v\do\w\do\x\do\y\do\z\do\A\do\B\do\C\do\D%
      \do\E\do\F\do\G\do\H\do\I\do\J\do\K\do\L\do\M\do\N%
      \do\O\do\P\do\Q\do\R\do\S\do\T\do\U\do\V\do\W\do\X%
      \do\Y\do\Z}
\def\UrlDigits{\do\1\do\2\do\3\do\4\do\5\do\6\do\7\do\8\do\9\do\0}
\g@addto@macro{\UrlBreaks}{\UrlOrds}
\g@addto@macro{\UrlBreaks}{\UrlAlphabet}
\g@addto@macro{\UrlBreaks}{\UrlDigits}
\makeatother

\interspeechcameraready

\title{TraceableSpeech: Towards Proactively Traceable Text-to-Speech with Watermarking}

\name[affiliation={1,2}]{Junzuo}{Zhou}
\name[affiliation={1, *}]{Jiangyan}{Yi}
\name[affiliation={1}]{Tao}{Wang}
\name[affiliation={3}]{Jianhua}{Tao}
\name[affiliation={1}]{Ye}{Bai}
\name[affiliation={1,2}]{Chu Yuan}{Zhang}
\name[affiliation={1,2}]{Yong}{Ren}
\name[affiliation={1}]{Zhengqi}{Wen}

\address{
  $^1$\small{Institute of Automation, Chinese Academy of Sciences, China
  $^2$School of Artificial Intelligence, University of Chinese  Academy of Sciences, China
  $^3$Department of Automation, Tsinghua University, China}}
\email{zhoujunzuo2023@ia.ac.cn, jiangyan.yi@nlpr.ia.ac.cn}


\begin{document}

\maketitle


\begin{abstract}
    
    
    Various threats posed by the progress in text-to-speech (TTS) have prompted the need to reliably trace synthesized speech. However, contemporary approaches to this task involve adding watermarks to the audio separately after generation, a process that hurts both speech quality and watermark imperceptibility. In addition, these approaches are limited in robustness and flexibility. To address these problems, we propose TraceableSpeech, a novel TTS model that directly generates watermarked speech, improving watermark imperceptibility and speech quality. Furthermore, We design the frame-wise imprinting and extraction of watermarks, achieving higher robustness against resplicing attacks and temporal flexibility in operation. Experimental results show that TraceableSpeech outperforms the strong baseline where VALL-E or HiFicodec individually uses WavMark in watermark imperceptibility, speech quality and resilience against resplicing attacks. It also can apply to speech of various durations.

\end{abstract}

\section{Introduction}

Recently, language model technology has achieved excellent performance in the text-to-speech (TTS) tasks such as VALL-E~\cite{VALLE}, SPEAR-TTS~\cite{speartts}, and SoundStorm~\cite{soundstorm}. These methods usually use neural codec~\cite{encodec,hificodec} to extract discrete representation from waveform and put them into language models for training.
Synthetic speech becomes increasingly realistic and natural, raising social issues regarding security and privacy, such as deepfake audio scams and copyright protection.
Therefore, it is vital for regulatory agencies supervise synthetic speech through traceability methods~\cite{baigong,china}.
Passive forensics is one of the most common options for traceability~\cite{yan1,yan2,yan3}. However, the artifact based detection are difficult to generalize well to unknown scenarios, making it susceptible to failure as the advancement of increasingly lifelike speech forgery techniques.

Based on the analysis above, proactive traceability in TTS systems is imperative. Responding to this necessity, several attempts of embedding watermarking signals as source information in the generated speech, aim to alleviate this problem. 
By utilizing specific algorithms to extract watermarks imperceptible to the ear, it is feasible to identify the source of the speech.

Audio watermarking methods are divided into two categories: traditional and deep learning based.
Traditional methods mainly include echo hiding~\cite{echo}, patchwork~\cite{2003patchwork}, spread spectrum~\cite{spreadspectrum}, etc. These methods have fragility and limited adaptability because they rely on expert knowledge and empirical rules. 
Meanwhile, increasingly powerful deep learning based frameworks can automatically model more robust watermark encoding via neural networks in a data-driven manner.
This advantage simplifies the watermarking design while keeping superior extractability against real-life speech manipulations or attacks.
Several works have been proposed based on the DNN network~\cite{pavlovic2022robust,Deorustc}. 
Recently, Chen et al.~\cite{2023wavmark} proposed the WavMark, an audio watermarking framework based on reversible networks, which surpasses previous work in each aspect.

However, embedding watermarks into generated speech through the above frameworks to achieve proactive traceability in TTS still has some limitations.
Firstly, watermark insertion is constrained to post-generation phases, which triggers error accumulation, reducing the watermark imperceptibility and the speech quality; 
Secondly, some advanced approaches (e.g. WavMark) exhibit issues of low temporal flexibility in implementation and suboptimal robustness against resplicing attacks.
Specifically, during inference, WavMark is restricted to embedding watermarks in speech segments that equate to the training snippets in duration, making it unsuitable for TTS tasks with unpredictable speech durations.
In addition, WavMark repeatedly embeds uniform watermarks on segments at fixed intervals to resist temporal edits. However, it is still susceptible to high-intensity resplicing attacks, particularly in shorter utterances.

To address these issues, we propose a proactively traceable TTS model named TraceableSpeech.
Firstly, for imperceptibility, TraceableSpeech integrates watermarking technology with language model based TTS via end-to-end training of codec and watermarking mechanism. 
It directly generates watermarked speech as information is embedded in the synthesis phase, optimizing watermark imperceptibility and speech quality; 
Secondly, for robustness and flexibility, we design a method for the frame-wise imprinting and extraction of watermarks. 
This method broadcasts watermark embedding and merges it with speech intermediate features of codec at frame-level and restore watermark from an \textit{r-vector} extracted by ResNet~\cite{resnet34_sp_re}, 
which maintains exceptional resilience under resplicing attacks and ensures availability across speech of various durations.
We conducted experiments on the LibriTTS~\cite{libritts}.
The contributions of this paper are as follows:
\begin{itemize}
\item We propose a proactively traceable TTS model jointly optimized for watermarking. This work generate watermarked speech directly, 
enhancing watermark imperceptibility to human listeners and speech quality as shown in better score on PESQ~\cite{pesq}, ViSQOL~\cite{visqol}, and subjective metrics, etc.

\item We design a watermark embedding and extraction method tailored to TTS tasks, ensuring the watermark’s robustness as shown in the better extraction accuracy after resplicing attacks, while offering temporal flexibility in operation. 
Even after embedding 4-digit base-64 watermarks in 0.3-second utterances, the extraction accuracy still remains above 95\%.
\end{itemize}

\begin{figure*}[t]
  \begin{minipage}[a]{0.79\linewidth}
    \centering
    \centerline{\includegraphics[width=\linewidth]{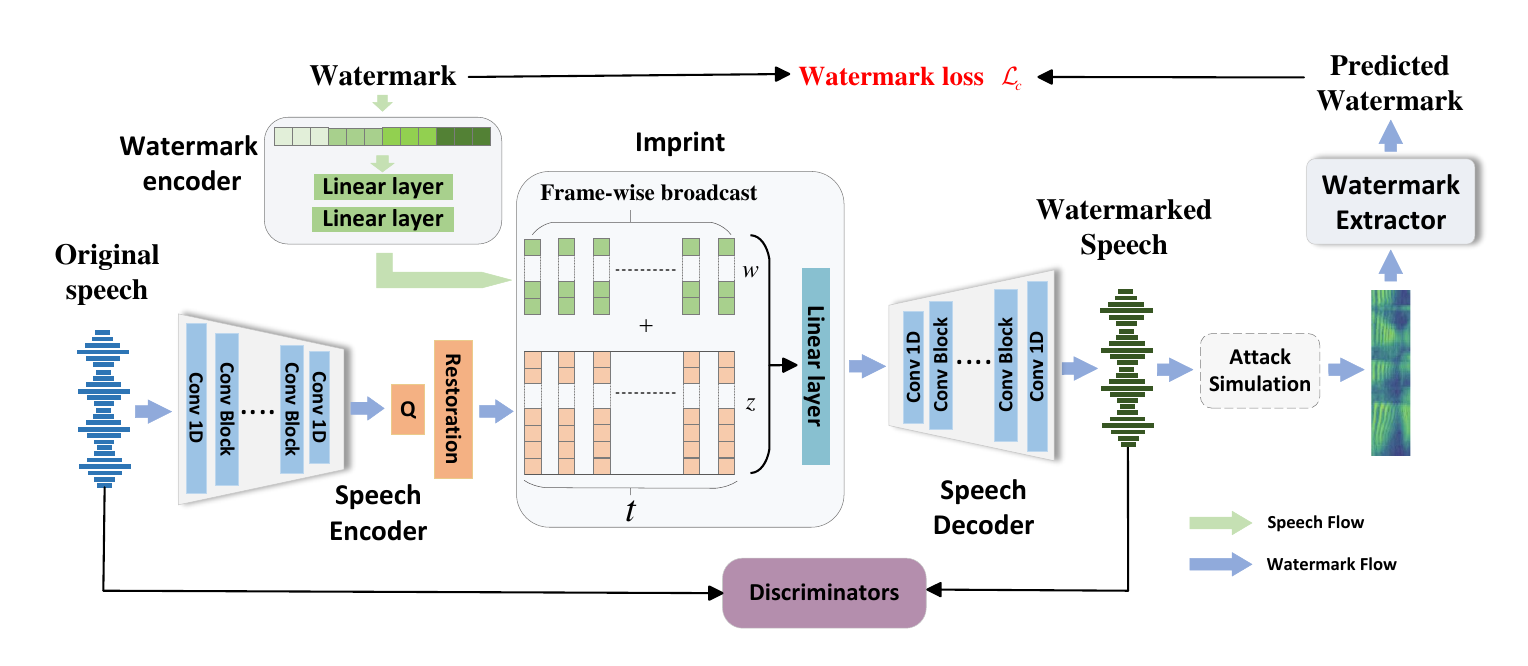}}
    \centerline{(a) Overall architecture.}\medskip
  \end{minipage}
  \hfill
  \begin{minipage}[a]{0.2\linewidth}
    \centering
    \vspace{0.4cm}
    \centerline{\includegraphics[width=\linewidth]{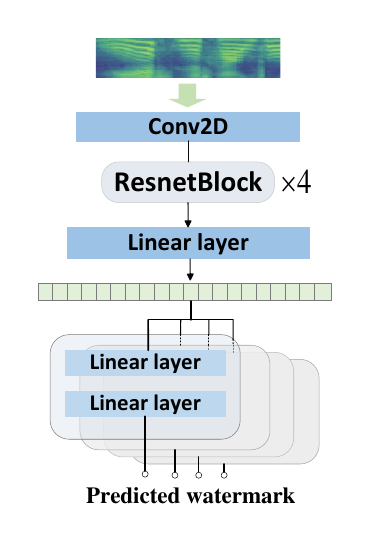}}
    \vspace{0.4cm}
    \centerline{(b) Watermark Extractor module.}\medskip
  \end{minipage}
  \caption{The first stage: Watermarking mechanism integrate into neural codec.}
  \label{fig: codec}
\end{figure*}

\begin{figure}[t]
  \centering
  \centerline{\includegraphics[width=1.12\linewidth]{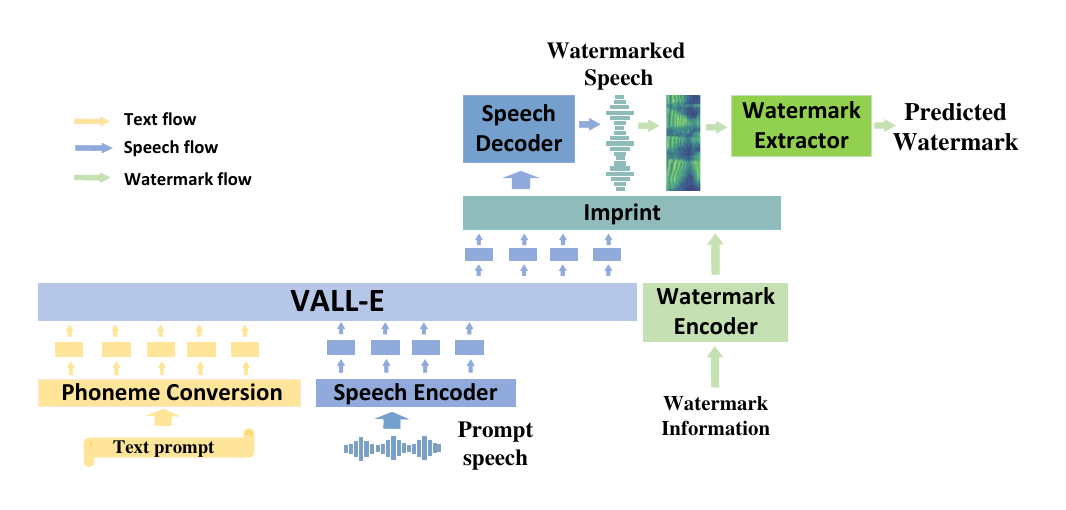}}
  \caption{The second stage: Watermarking mechanism integrate into language model of VALL-E.}
  \label{fig: VALL-E}
\end{figure}

\section{Proposed Method}

\subsection{The overall Framework of TraceableSpeech}

The speech synthesis in TraceableSpeech is structured into two sequential stages: the neural codec and the language model. Figures \ref{fig: codec} and \ref{fig: VALL-E} respectively illustrate the integration of the watermarking mechanism into these two stages. Both stages realize the closed-loop process of information embedding to retrieval via modules such as watermark encoder, imprint, and watermark extractor.

As shown in Figure \ref{fig: codec}, the neural codec utilizes speech encoder and speech decoder both derived from HifiCodec’s design~\cite{hificodec}. 
The speech waveform of duration $d$ is represent as $x \in \mathbb{R}^T$ with a sampling rate of $f_{sr}$, where $T=f_{sr}\times d$. 
In training, the speech waveform undergoes encoder downsampling, quantization, and quantized restoration, thereby transforming into a high-dimensional latent representation $z \in \mathbb{R}^{t\times512}$, where $t$ is the count of frames after 240$\times$ down-sampling. 
The watermark information is embedded in $z$ through the imprint module. Then, the speech decoder generates watermarked speech from~$z$. Ultimately, the joint end-to-end training of watermarking and codec is realized utilizing the watermark decoder and discriminators.

As shown in Figure \ref{fig: VALL-E}, the discrete representation obtained by the speech encoder is put into the language model with the same structure as VALL-E.
During inference, the imprint module embeds the watermark information into the discrete representation predicted from the language model. Then, the watermarked speech is synthesized by the speech decoder.

\subsection{Frame-wise Imprinting}

Previous methods directly extend watermark vector through linear layers to match the waveform's length~\cite{2023wavmark}, sacrificing temporal flexibility and raising non-uniform distribution of watermark information along the temporal axis. 
In this work, watermark information is embedded into frame-level speech features and control it by broadcasting in time-domain, thereby supporting speech of various duration.
Furthermore, the embedded information is uniform and comprehensive across all parts of this speech, which avoids damage from resplicing attacks.

As shown in Figure \ref{fig: codec}(a), we utilize $m$-digit base-$b$ numerical information as a watermark. The watermark encoder first maps the number of each digit to embedding using a $b\times16$ weight matrix. Then, two linear layers convert a long vector concatenated by $m$ embedding into latent representation $w_{o} \in \mathbb{R}^{1\times512}$.
In the imprint module, $w_{o}$ is broadcast along the time axis. Therefore, by controlling the position and number of frames that are broadcast, the positions and duration of the watermarked segments can be precisely controlled in the synthesized speech.

In practice, to ensure full-time region protection. All frames are broadcast to obtain $w \in \mathbb{R}^{t\times256}$ that is the high-dimensional feature for merging with the speech latent representation $z$. 
So that, even if some frames are truncated, the remaining frames can still be successfully extracted. 
Since the watermark is embedded at the frame level, the broadcast imprint module can embed the information into the entire speech of any duration, thereby achieving broad temporal flexibility.

\subsection{Watermark extractor and training mechanism}

\subsubsection{Watermark extractor}

As shown in Figure \ref{fig: codec}(b), the input of the watermark extractor is the Mel-spectrogram of the synthesized speech from the speech decoder. 
An \textit{r-vector} extracted through the ResNet~\cite{wespeaker} is individually connected with $m$ groups of two-layer linear layers to calculate the probability distribution of each digit and obtain the predicted number by softmax. 

\subsubsection{Training mechanism}
\textbf{Attack simulation}: We incorporate attack simulation in training to 
be resilient against common watermark attack. In this module, the synthesized speech undergoes one of the following seven processes~\cite{Deorustc,2023wavmark}: 
Normal extraction, no attack (Normal);
Resample with 90\% of original sampling rate and recovery (RS-90);
White noise with an SNR of 35db was added (Noise-W35);
Randomly dropout 0.1\% of the sample points~(SD-01);
Reduce the amplitude to 90\% of its original value~(AR-90);
Attenuate the resulting speech by a factor of 0.3, delay the volume by 15\%, then overlay the original speech~(EA-0315);
Low-pass filter with a cutoff frequency of 5k Hz. Since high-pass filtering would destroy the essential information for synthesized speech, only the more practical low-pass filtering is considered~(LP-5000).

The assigned weight values of the aforementioned processes are empirically set at 0.45, 0.04, 0.25, 0.04, 0.04, 0.14, and 0.04, respectively, due to the TraceableSpeech's higher sensitivity to Noise and Echo.

\noindent \textbf{Optimizing strategy}: To integrate with the above network, we design the following optimization strategy. The training process is divided into two stages: neural codec and language model. Compared with HiFiCodec~\cite{hificodec}, the loss function of the neural codec adds the cross-entropy watermark loss in the generator:
\begin{align}
    \mathcal{L}_c=\frac{1}{m} \sum_{i=1}^{m} \sum_{j=1}^{b} l_{ij} \log \left(p_{i j}\right)
\end{align}
where $l_{ij}$ and $p_{ij}$ are the one-hot encoding and the predicted probability of the $i$-th digit watermark for number $j$, respectively.
In addition, the generator loss also include 
the reconstruction loss of frequency domain $\mathcal{L}_{f}$, 
the quantization loss $\mathcal{L}_{qz}$, 
the feature matching loss $ \mathcal{L}_{feat}$, 
and the adversarial loss of the generator $\mathcal{L}_{g}$, 
all of which are the same as HiFiCodec. The total loss function of the generator is:
\begin{equation}
\begin{split}
  \mathcal{L} & = \lambda_f \mathcal{L}_f + \lambda_g \mathcal{L}_g + \lambda_{feat} \mathcal{L}_{feat} + \lambda_{qz} \mathcal{L}_{qz} + \lambda_{c} \mathcal{L}_c
\end{split}
\end{equation}

$\lambda_f$, $\lambda_g$, $\lambda_{feat}$, $\lambda_{qz}$ and $\lambda_{c}$ are hyper-parameters to balance each term of the final loss. Their values are 1, 1, 1, 10, and 5, respectively, thus reflecting a bias towards the watermark network and quantizer.

The training of the language model is the same as VALL-E.

\section{Experiments}

\subsection{Dateset}

We use the LibriTTS dataset to train the neural codec\footnote[1]{https://github.com/yangdongchao/AcademiCodec} and the VALL-E\footnote[2]{https://github.com/lifeiteng/vall-e} language model from scratch. The LibriTTS corpus~\cite{libritts} consists of 585 hours of English speech data from 2456 speakers at 24kHz. Our training set consists of train-clean-100, train-clean-360, and train-other-500. Our test set is also from the subsets of LibriTTS.

\subsection{Experiment Setup}

\noindent \textbf{Baseline}: We compare TraceableSpeech with the state-of-the-art deep audio watermarking framework\footnote[3]{https://github.com/wavmark/wavmark}~\cite{2023wavmark}. This framework is trained on the 1-second audio snippet.
Hence, watermarking can only be applied to 1-second audio segments during inference.
It utilizes an “utterance mode” for audio exceeding this length by repeatedly adding the same 1-second watermark content at fixed intervals.
While WavMark embeds 32-bit binary watermarks in this mode, 
the initial 16 bits are allocated as pattern bits to ascertain the validity and completeness of this segment's watermark.
This method notably reduces the usable capacity to 16 bits in binary. This watermark is deemed unextractable if the pattern bits in all added segments are identified as failures. 
In speech reconstruction and  zero-shot speech synthesis, we utilize WavMark to embed watermarks into the speech waveforms generated by HiFicodec and VALL-E, respectively. These watermarked speech are used for comparison.

\noindent \textbf{Training setup}: 
In the experiments detailed in section \ref{Speech Reconstruction} and \ref{Zero-Shot}, we train a 4-digit base-16 model $4@16$, which has the same watermark capacity as the baseline, and a 4-digit base-10 model $4@10$.
We use ResNet34 in the watermark extractor. The dimension of the extractive embedding is 256.
For the neural codec, the quantizer utilizes 1 group with 8 codebooks and the batch size is 32. We truncate the training data to 0.5 seconds, all models are trained for 150k steps.
For the language model, the maximum duration per batch is 100. The AR and NAR stages are trained for 20 and 40 epochs, respectively.

\noindent \textbf{Resplicing attacks setup}: During inference, a resplicing attack means that the watermarked speech is randomly cut out $1/4$ to $1/3$ of the watermarked speech is cut out from the middle of the original waveform, with the rest concatenated.


\begin{table}[t]
  \centering
  \caption{Watermark Imperceptibility Metrics in Speech Reconstruction}
  \label{tab:first}
  \scalebox{0.90}{
  \begin{threeparttable}
  \begin{tabular}{lccc}
    \toprule
    \textbf{Model}   & \textbf{PESQ $\uparrow$} & \textbf{STOI $\uparrow$} & \textbf{ViSQOL $\uparrow$} \\
    \midrule
    HiFicodec + WavMark(16bit)   & 3.197                    & 0.947                         & 3.880                      \\
    \midrule
    TraceableSpeech(4@10) & \textbf{3.641}                    & \textbf{0.950}                    & \textbf{4.060}                      \\
    TraceableSpeech(4@16) & 3.569                    & 0.948               & 3.985                      \\
    \bottomrule
  \end{tabular} 
  \begin{tablenotes}    
        \footnotesize               
        \item[1] $@$ denotes the watermarking capacity. For example, $4@16$ indicates 4-digit base-16, equivalent to the 16-bit capacity of WavMark used in the baseline. This annotation is applicable to other tables as well.
    \end{tablenotes}            
  \end{threeparttable}
  }
\end{table}

\begin{table}[t]
  \caption{Speech Quality in Zero-Shot Speech Synthesis}
  \label{tab:second}
  \centering
  \scalebox{0.9}{
  \begin{tabular}{lcc}
    \toprule
    \textbf{Model}        & \textbf{WER$(\%)$ $\downarrow$}  & \textbf{MOS $\uparrow$} \\
    \midrule
    VALL-E + WavMark(16bit)       &  10.80                           & 3.554 $\pm$ 0.19           \\
    \midrule
    TraceableSpeech(4@10) & \textbf{9.61}                    & \textbf{3.959} $\pm$ 0.18  \\
    TraceableSpeech(4@16) & 10.47                            & 3.905 $\pm$ 0.17           \\
    \bottomrule
  \end{tabular} 
  }
\end{table}

\subsection{Performance of Speech Reconstruction} \label{Speech Reconstruction}

Comparing the watermarked speech with its unwatermarked counterpart can help evaluate the watermark imperceptibility, achieved by calculating PESQ~\cite{pesq}, STOI~\cite{stoi}, ViSQOL V3~\cite{visqol} metrics.
Since the speech generated in speech synthesis experiment is still diverse even using the same text, it is necessary to set up a codec speech reconstruction experiment. 
200 test speech samples of various durations are from the test-clean of the LibriTTS corpus
Each metric is calculated by comparing the reconstructed speech with the ground truth. 
Table \ref{tab:first} demonstrates that TraceableSpeech outperforms baselines in all metrics. Additionally, The comparison of 4@10 and 4@16 indicates that the watermark imperceptibility diminishes as its capacity increases.

\begin{table*}[t]
  \caption{Watermark extraction accuracy (\%) under various attacks}
  \label{tab:third}
  \centering
  \scalebox{1.0}{
  \begin{threeparttable}
  \begin{tabular}{lcccccccc}
    \toprule
    \diagbox{\textbf{Model}}{\textbf{Attack}}  & \textbf{Resplicing} & \fontsize{10pt}{\baselineskip}\selectfont\textbf{Normal} & \textbf{RSP-90}  & \textbf{Noise-W35}  & \textbf{SD-01} & \textbf{AR-90} & \textbf{EA-0315}  & \textbf{LP5000}   \\
    \midrule
    VALL-E + WavMark(16bit)       & No          & 100.00                   & 99.76                    & 91.41            & 100.00         & 100.00     & 94.53                 & 100.00       \\
    TraceableSpeech(4@10) & No          & 100.00                   & \textbf{100.00}          & \textbf{100.00}  & 100.00         & 100.00     & \textbf{100.00}       & 100.00       \\
    TraceableSpeech(4@16) & No          & 98.97                    & 98.82                    & 98.95            & 99.12          & 99.46      & 97.71                 & 98.84        \\
    \midrule
    VALL-E + WavMark(16bit)              & Once        & 91.10                    & 91.46                    & 63.53            & 95.95          & 93.61           & 88.58            & 89.66         \\
    TraceableSpeech(4@10) & Once        & \textbf{100.00}          & \textbf{100.00}          & \textbf{100.00}  & \textbf{99.90} & \textbf{100.00} & \textbf{100.00}  & \textbf{100.00} \\
    TraceableSpeech(4@16) & Once        & 100.00                   & 99.82                    & 99.83            & 98.78          & 99.50           & 99.57            & 99.62          \\
    \midrule
    VALL-E + WavMark(16bit)       & Twice                    & 76.65                    & 77.74                    & 49.14      & 79.47                    & 85.46    & 68.19                    & 75.32               \\
    TraceableSpeech(4@10) & Twice         & \textbf{100.00}                    & \textbf{100.00}                    & \textbf{100.00}      & \textbf{100.00}                    & \textbf{100.00}    & \textbf{100.00}                    & \textbf{100.00}               \\
    TraceableSpeech(4@16) & Twice                   & 99.58                    & 99.20                    & 99.58      & 99.56                    & 99.00    & 99.65                    & 98.83               \\
    \bottomrule
  \end{tabular} 
    \begin{tablenotes}    
        \footnotesize               
        \item[1] The resplicing column mean the times of resplicing attack
    \end{tablenotes}            
  \end{threeparttable}
  }
\end{table*}

\begin{table*}[h]
  \caption{Watermark extraction accuracy (\%) of larger capacity models under various speech durations (s)}
  \label{tab:fourth}
  \centering
  \scalebox{1.0}{
  \begin{tabular}{lccccccccc}
    \toprule
    \diagbox{\textbf{Model}}{\textbf{Duration}}   & \textbf{1.0} & \textbf{0.8} & \textbf{0.5}   & \textbf{0.3} & \textbf{0.2} & \textbf{0.175} & \textbf{0.15} & \textbf{0.125} & \textbf{0.1}   \\
    \midrule
    TraceableSpeech(4@32) & 100.00        & 100.00          & 99.74          & 99.23        & 94.13        & 86.22          & 77.29         & 57.14          & 50.51          \\
    TraceableSpeech(4@64) & 100.00        & 100.00          & 99.86          & 95.57        & 80.59        & 66.79          & 53.90         & 27.47          & 17.01          \\
    \bottomrule
  \end{tabular} 
  }
\end{table*}

\subsection{Performance of Zero-Shot Speech Synthesis} \label{Zero-Shot}

We use 200 text prompts from the test-clean of the LibriTTS corpus. 
Each sample is subjected to 20 tests of watermark embedding and extraction.
The duration of the synthesized speech is restricted between 1.125 seconds and 10 seconds to reflect the temporal diversity. Considering the limit of the baseline, we also exclude test samples that are shorter than the aforementioned lower bound after resplicing attacks.

The quality of the synthesized watermarked speech can be evaluated using subjective and objective metrics.
We utilized HuBERT-large-ls960-ft4\footnote[5]{https://huggingface.co/facebook/hubert-large-ls960-ft}~\cite{hubert} to transcribe speech and compute WER to evaluate content accuracy. 
In addition, We invited seven participants to mark speech quality with MOS results. 
Table \ref{tab:second} shows the results of speech quality, with our work surpasses baselines. And the speech quality also exhibits an inverse relationship with the watermark capacity.

If the watermark in the baseline cannot be extracted, it is considered that all bits are incorrect.
As shown in Table \ref{tab:third}, the robustness results indicate that our work maintains a higher extraction accuracy when facing resplicing attacks than the baseline. Furthermore, our advantage becomes increasingly apparent as the attacks intensify.

\subsection{Quantitative analysis of Capacity and Duration}

The analysis explores the impact of increased watermark capacity and reduced speech duration on extraction accuracy.
Because the robustness, capacity, and imperceptibility of watermarks are impossible to achieve simultaneously, 
the models trained in this analysis, including 4-digit base-32 (4@32) and 4-digit base-64 (4@64), are not subjected to simulated attacks.
Considering the increased capacity, we use ResNet101~\cite{resnet34_sp_re} in the watermark extractor, and the dimension of the extractive embedding is 512.
This analysis is conducted through speech reconstruction to precisely control the duration of the speech for evaluation.
The test set is composed of speech slices ranging from 0.1 to 1 second.
As shown in Table 4, even after embedding 4-digit base-64 watermarks in 0.3-second speech segments, the extraction accuracy of our method still remains above 95\%.

\section{Conclusion}

This work proposes TraceableSpeech, a novel TTS model that jointly optimizes the watermarking mechanism and speech synthesis, thereby directly generating watermarked speech. This approach enhances the watermark imperceptibility and speech quality.
This work also proposes frame-wise imprinting and extraction networks of watermarks, designed specifically for the characteristics of the TTS task to enhance robustness against resplicing attacks and improve temporal flexibility for speech of various durations. 
Experimental results demonstrate that TraceableTTS performs superiorly in various metrics, including PESQ, WER, and extraction accuracy after  resplicing attacks.
In the future, We aim to bolster the robustness against increasingly varied and more potent watermark attacks. Finally, The code is avaliable at \href{https://github.com/zjzser/TraceableSpeech}{\textcolor{blue}{https://github.com/zjzser/TraceableSpeech}}

\section{Acknowledgements}

This work is supported by the Strategic Priority Research Program of Chinese Academy of Sciences, Grant No. XDB0500103, the National Natural Science Foundation of China (NSFC) (No. 62322120, No.U21B2010, No. 62306316, No. 62206278).

\bibliographystyle{IEEEtran}
\bibliography{mybib}

\end{document}